\documentclass[aps,notitlepage,nofootinbib]{revtex4-1}

\usepackage{amstext,amsmath,amssymb,amsfonts,bbm}
\usepackage[latin1]{inputenc}
\usepackage{epsfig}
\usepackage{hyperref}
\usepackage{amsthm}
\usepackage{subfigure}
\usepackage{color}
\usepackage{multirow}

\newcommand{\bea}{\begin{eqnarray}}
\newcommand{\eea}{\end{eqnarray}}
\newcommand{\be}{\begin{equation}}
\newcommand{\ee}{\end{equation}}
\newcommand{\cG}{{\cal G}}
\newcommand{\cR}{{\cal R}}

\newcommand{\cT}{{\cal T}}

\newcommand{\cF}{{\cal F}}
\newcommand{\cB}{{\cal B}}
\newcommand{\cL}{{\cal L}}
\newcommand{\cN}{{\cal N}}
\newcommand{\cJ}{{\cal J}}
\newcommand{\cH}{{\cal H}}
\newcommand{\cS}{{\cal S}}

\newtheorem{lemma}{Lemma}

\newtheorem{definition}{Definition}

\begin{document}

\title{The complete $1/N$ expansion of colored tensor models in arbitrary dimension}

\author{Razvan Gurau}\email{rgurau@perimeterinstitute.ca}

\affiliation{Perimeter Institute for Theoretical Physics, 31 Caroline
St, Waterloo, ON, Canada}

\begin{abstract}
\noindent In this paper we generalize the results of \cite{colorN,colorNleadD} and derive the full $1/N$ expansion 
    of colored tensor models in arbitrary dimensions.
    We detail the expansion for the independent identically distributed model and the 
    topological Boulatov Ooguri model.
\end{abstract}

\maketitle

\section{Introduction}

Random matrix models \cite{mm} generalize in higher dimensions to 
random tensor models \cite{mmgravity,ambj3dqg,sasa1} and Group Field 
Theories (GFT) \cite{Boul,Ooguri:1992eb,laurentgft,quantugeom2}. 
The Feynman graphs of GFT in $D$ dimensions are built from vertices dual to 
$D$ simplices and propagators encoding the gluing of simplices along their boundary. 
Parallel to ribbon graphs of matrix models (dual to discretized surfaces), GFT 
graphs are dual to discretized $D$ dimensional topological spaces.

The perturbative series of random matrix models can be reorganized in powers of $1/N$ (where $N$ is the dimension 
of the matrices). The leading order consists in planar graphs (paving the 2 dimensional sphere $S^2$) 
\cite{Brezin:1977sv} with higher genus graphs suppressed in powers of $N$. The $1/N$ series is crucial to the
continuum limit of random matrix models and is at the core of their applications to  
confinement \cite{'tHooft:1973jz}, string theory \cite{Gross:1990ay,Gross:1990ub}, 
two dimensional gravity \cite{Di Francesco:1993nw}, critical phenomena \cite{mm,Kazakov:1985ea,Boulatov:1986jd}, 
black hole physics \cite{Kazakov:2000pm}, etc. 

In the most common GFT models \cite{Boul,Ooguri:1992eb} the Feynman amplitude 
of a graph equals the partition function of BF theory 
\cite{FreidelLouapre,gftnoncom} discretized on the 
dual gluing of simplices. GFT is a fast evolving field, and progress has
been made in several directions. More involved models
\cite{newmo2, newmo3, newmo4, newmo5, gftquantgeom,gftnoncom} 
have been proposed in an attempt 
to implement the Plebanski constraints and reproduce the gravity partition function.
This models play in higher dimensions the same role non identically distributed matrix models \cite{GW,GW1} 
play in two dimensions. As some of the latter have been proven ultraviolet 
complete \cite{beta1,beta2} one can hope that a UV complete GFT can be built.
The semiclassical limit of various GFT models 
\cite{semicl1, semicl1.5} has been analyzed and matter field have been added in the GFT framework
\cite{matter0.1,matter0.2, matter0.3}. Alternatively, effective field theories
which admit an interpretation as non commutative matter fields have been derived
\cite{matter1,matter2} by developing the GFT action around particular instanton solutions.
The implementation of the diffeomorphism symmetry at the level of the GFT action \cite{Baratin:2011tg}
has been studied. Recently GFT's have been adapted to the study of loop quantum cosmology \cite{Ashtekar:2009dn,Ashtekar:2010ve}.
Partial power counting theorems and bounds on GFT amplitudes 
\cite{FreiGurOriti,sefu1,sefu2,sefu3} have been obtained, and the radiative corrections
of the Boulatov model derived \cite{Geloun:2011cy}.

In spite of these numerous developments, the question of the $1/N$ expansion of GFTs has gone a 
long time unanswered. Indeed, it has been identified as one of the most important challenges in 
the field \cite{Alexandrov:2010un}. 
It is tempting to conjecture that GFTs in $D$ dimensions should admit also a $1/N$ 
expansion in which dominant graphs pave the $D$ dimensional sphere $S^D$,
but no indication of this has been found until recently. 

A good deal of progress has been made for the newly introduced ``colored'' \cite{color,PolyColor,lost} GFT (CGFT)
models. The full $1/N$ expansion of CGFTs in $D=3$ dimensions has been performed \cite{colorN}, and 
subsequently the dominance at leading order of the spherical topology has been established in
arbitrary dimensions \cite{colorNleadD}. This paper constitutes the last piece of the puzzle:
building on the results of \cite{colorN,colorNleadD} we derive here the complete $1/N$ series in 
arbitrary dimension. 

We will perform the expansion for two models, the colored independent identically distributed (i.i.d.) model
and the colored Boulatov Ooguri \cite{Boul,Ooguri:1992eb} model in arbitrary dimensions. In higher dimensions
one needs to deal with a number of difficulties. The $1/N$ expansion in matrix models (which we will review in detail 
in section \ref{sec:mm}), is indexed by the genus. The genus is the unique topological invariant in two dimensions.
This leads to an expansion which is at the same time an amplitude expansion {\it and} a topological expansion.
In higher dimensions one does not have the luxury of such an invariant. Indeed, although the amplitude expansion
captures some topological aspects (for instance it is dominated by graphs with spherical topology \cite{colorNleadD}),
it is {\it not} a priori a topological expansion. 

We present in this paper two alternative expansions of the colored GFT models. First, we perform an expansion exclusively 
in the amplitude. We call this a ``combinatorial expansion'' and we will detail it at length for the i.i.d. model. This
expansion is canonical, but completely ignores the topology. We will subsequently present a second expansion taking into 
account not only the amplitude but also also the topology of graphs. This requires a good 
deal of extra work as one needs to classify graphs of the same amplitude into topological classes. 
The classification we perform in this paper is neither canonical nor unique. It is however simple enough such that it leads to 
a workable series one can evaluate term by term. We will detail this ``topological expansion'' both for the i.i.d. model
and for the Boulatov Ooguri model.

We would like to emphasize that none of the concepts and techniques needed to perform the $1/N$ expansion
can directly be applied to the non colored GFT models. This leads us to believe that a similar expansion 
either does not exist in the ordinary models, or it has a very different nature from the one of
CGFTs.

This paper is a complete and self contained presentation of the techniques required to perform the 1/N 
expansion in arbitrary dimensions. In section \ref{sec:mm} we revisit
the $1/N$ expansion of random matrix models.
Section \ref{sec:graphs} introduces the CGFT graphs and discusses the associated notions of bubbles and jackets.
Section \ref{sec:models} describes briefly the generic colored GFT  models, and in sections
\ref{sec:iid} and \ref{sec:topomodel} we perform the $1/N$ expansion of CGFT.
Section \ref{sec:firstterms} presents the first terms in our expansion in arbitrary dimensions.

\section{The $1/N$ expansion of random matrix models}\label{sec:mm}

In this preliminary section we will recast the familiar $1/N$ expansion of matrix models in 
a form adapted to the generalization in higher dimensions. The partition function of an 
i.i.d. $N \times N $ matrix model writes
\bea\label{eq:rib}
  e^{-F(\lambda)}=Z(\lambda) = \int dM e^{-N \Big{(}\frac{1}{2} \text{Tr} M^2 +  \sum_p \lambda_p \text{Tr} M^p \Big{)}} \; ,
\eea
and the free energy $F$ develops in vacuum ribbon graphs $\cG$ with $\cN_{\cG}$ vertices, $\cL_{\cG}$ lines, 
$\cF_{\cG}$ faces, and genus $2-2g_{\cG}= \cN_{\cG} - \cL_{\cG}+\cF_{\cG}$. 
For the i.i.d. model one can directly compute the amplitude of any graph, $A(\cG) = N^{\cN_{\cG}-\cL_{\cG}+ \cF_{\cG} }$,
and the free energy admits an expansion in the genus. It reads (for a single value of $p$ and $\lambda = \lambda_p$)
\bea\label{eq:matrixgenus}
 F(\lambda) = \sum_{g=0}^{\infty} C^{[g]}(\lambda)
 N^{2 - 2g} \; ,\qquad
C^{[g]}(\lambda) = \sum_{\cG, \; g_{\cG} = g } \frac{ 1 } {s(\cG)} \lambda^{\cN_{\cG}}  \; ,
\eea 
where $C^{[g]}(\lambda)$ is the sum of the series of graphs of fixed genus $g$, and $s(\cG)$
a symmetry factor. As the genus
is the only topological invariant in $D=2$ dimensions, the expansion in $1/N$ is a topological expansion.
The series $C^{[g]}(\lambda)$ can be analyzed by 
a multitude of techniques \cite{Am1,Am2,diageq} and encodes the critical behavior of the model.

For non identically distributed matrix models \cite{GW,GW1} one can not directly compute the amplitude
of an arbitrary Feynman graph. One accesses its scaling with $N$ by an alternative method of 
topological reduction moves. We present here this second method on i.i.d. models, but the reader should 
keep in mind that this generalizes to arbitrary matrix models.
Note that the vertices and the faces bring each a factor $N$, and the lines bring a factor $N^{-1}$ to 
the amplitude of a graph. There exist two local transformations of 
$\cG$ which leave both its amplitude and topology invariant. Under these transformations,
$\cG$ will simplify to some standard form which we subsequently analyze.

\begin{figure}[htb]
\begin{center}
 \includegraphics[width=3cm]{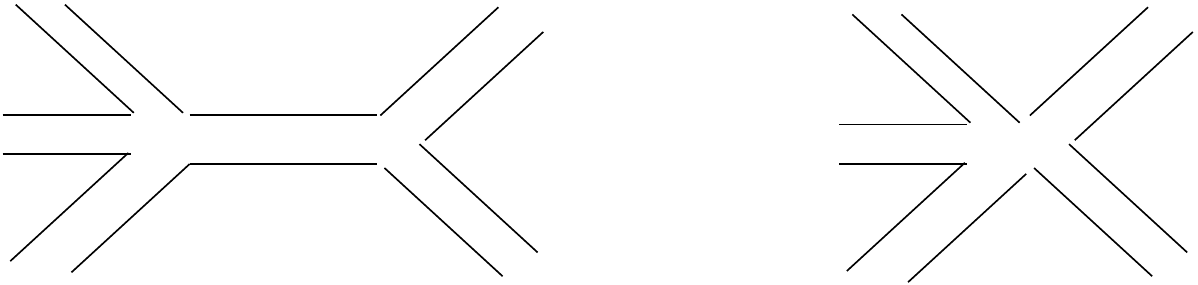}  
\caption{Contraction of a ribbon line in $\cG$. }
\label{fig:trecont}
\end{center}
\end{figure}

The first simplification move involves a line $l$ connecting two distinct vertices in the graph $\cG$. Any such line
can be ``contracted'' and the two adjacent vertices glued coherently, as drawn in figure \ref{fig:trecont}. 
We denote $\cG/l$ the graph obtained by contracting $l$. This move preserves the topology: 
$\cG/l$ has one less vertex and one less line than $\cG$, hence $g_{\cG/l} =g_{\cG}$. 
Unsurprisingly it also preserves the amplitude, $ A(\cG/l) = N^{-1} N A(\cG) $.
 
\begin{figure}[htb]
\begin{center}
 \includegraphics[width=4cm]{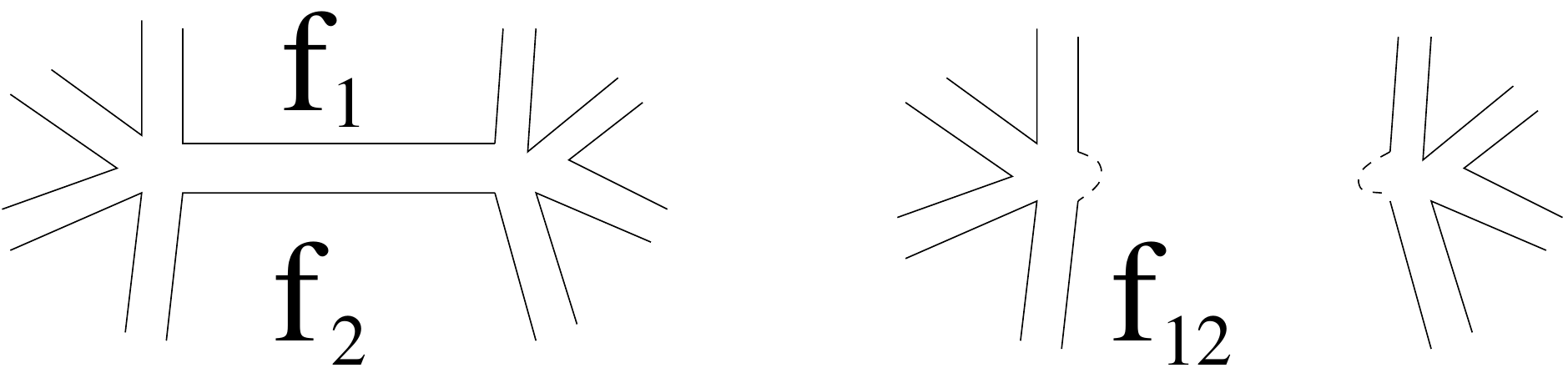}
\caption{Deletion of a ribbon line in $\cG$.}
\label{fig:del}
\end{center}
\end{figure}

The second simplification move involves a line $l$ separating two different faces. Such a line can be
``deleted'' and the two faces $f_1$ and $f_2$ merged into a unique face $f_{12}=f_1\cup f_2$ 
(see figure \ref{fig:del}).
We denote $\cG-l$ the graph obtained by deleting $l$. The deletion also preserves the topology ($\cG-l$ has one 
less line and one less face than $\cG$) and the amplitude.

Any two graphs $\cG$ and $\cG'$ related by a sequence of deletion/contraction moves (and their inverse) 
have the same topology and the same amplitude. Two such graphs are called ``equivalent'' $\cG \sim \cG'$. 
A graph $\cG$ can be maximally simplified using this moves 
to a ``super rosette'' graph \cite{param} $\cG_p\sim \cG$, having a unique vertex, a unique face and $2p$ lines. 
The super rosette to
which $\cG$ reduces is not unique (it depends on the specific lines one contracts and deletes), 
but all are equivalent $\cG_p'\sim \cG_p$. Furthermore, as
the number of contraction (resp. deletion) moves the graph undergoes to become a super rosette is 
$\cN_{\cG}-1$ (resp. $\cF_{\cG}-1$), the number of lines of all super rosettes 
corresponding to $\cG$ is $ 2p = \cL_{\cG}- (\cN_{\cG}-1) - (\cF_{\cG}-1)$. All super rosettes
with the same number of lines are equivalent and have amplitude $A(\cG_p)= N^{2-2p }$.
We divide the super rosettes in equivalence classes $[\cG_p]$, under the relation $\sim$, and 
write 
\bea\label{eq:matrixroset}
 F = \sum_{[\cG_p]} A ([\cG_p]) \; C^{[\cG_p]}(\lambda) = \sum_{p } N^{2-2p} \; C^{[\cG_p]}(\lambda) \; ,
\eea
where $A ([\cG_p])$ is the amplitude of the equivalence class of super rosettes $[\cG_p]$ and 
$C^{[\cG_p]}(\lambda)$ is a purely combinatorial coefficient counting all graphs which reduce to $[\cG_p]$. 
The two expansions, the genus expansion of eq. \eqref{eq:matrixgenus} and the rosette expansion 
of eq. \eqref{eq:matrixroset} are of course one and the same as $2p = 2g_{\cG_p}=2g_{\cG}$.

We will generalize these two expansions in higher dimensions. We will first derive in 
arbitrary dimensions an expansion in the ``degree'' generalizing directly eq. \eqref{eq:matrixgenus}
for i.i.d. models. We will then identify a set of moves which leave 
invariant both the topology and the amplitude of a graph. We will apply the moves to any graph to reduce 
it to some standard simplified graph (core graph). We will reorganize
the free energy into core graph classes, and bound the amplitude of every class. 
At any order we will obtain contributions from a finite number of topologies, and control over
the scaling of the remainder terms.

The topology of higher dimensional pseudo manifolds is more involved than the one of two dimensional 
surfaces, and the $1/N$ expansion will accordingly have some new and more complicated features.
\begin{itemize}
 \item For all $D\ge 3$ there exist moves which preserve the topology but bring power counting convergence.
   Unlike in matrix models, the same topology will be represented by an infinity of (increasingly suppressed)
   terms. 
 \item For all $D\ge 4$ establishing topological equivalence is very difficult. A classical mathematical 
   theorem implies that it can not be done by any algorithm and one must proceed 
   on a case by case basis \cite{Stillwell}.

\end{itemize}

In order to obtain a workable expansion we will only partially classify graphs to ensure 
that one needs to check topological equivalence a minimum number of times. 
The price to pay is that our final series has a (finite) degree of redundancy: graphs with the 
same topology and amplitude will be grouped in a (finite) number of distinct classes. 

\section{Graphs, Jackets and Bubbles}\label{sec:graphs}

In this section we prove some general results concerning CGFT graphs. The $1/N$ expansion 
relies largely on these results, and it is insensitive to the particulars of the model one deals with. 

\begin{figure}[htb]
\begin{center}
 \includegraphics[width=2cm]{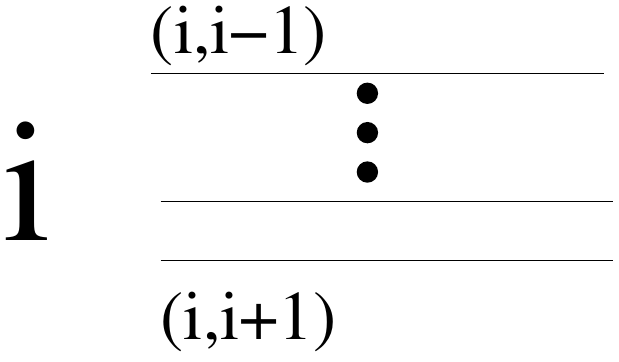} \hspace{1cm}
 \includegraphics[width=2cm]{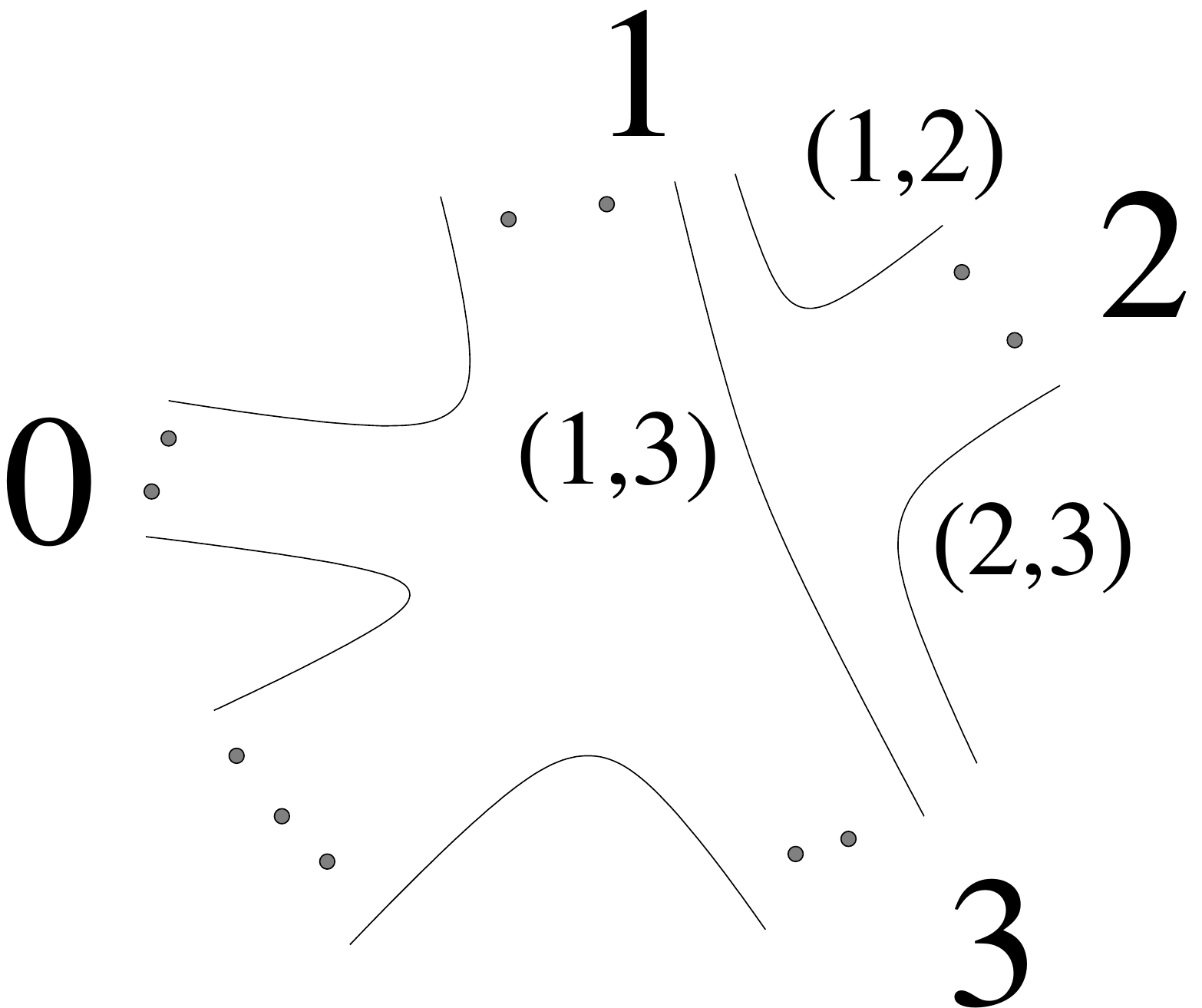} 
\caption{Line and vertex of the Colored GFT graphs.}
\label{fig:propvert}
\end{center}
\end{figure}

The Feynman graphs of the D-dimensional colored GFT (called (D+1)- colored graphs), 
are made of oriented colored lines with $D$ parallel strands and oriented vertices (dual to $D$ simplices) of
coordination $D+1$. For every vertex, the strand $(i,j)$ connects the half lines of colors $i$ and $j$ 
(see figure \ref{fig:propvert}). The $2+1$-colored graphs are familiar ribbon graphs of matrix models.
We denote $\cN_{\cG},\; |\cN_{\cG}|=2p$, $\cL_{\cG}$, $\cF_{\cG}$, the sets of vertices, lines and  faces 
(i.e. closed strands) of $\cG$.

A CGFT graph admits a simplified representation obtained by collapsing all the stranded vertices and 
stranded lines into point vertices and ordinary colored lines.
Colored graphs \cite{color,lost,PolyColor,caravelli} are dual to oriented normal pseudo manifolds. 
In this paper we will only deal with connected graphs.

\begin{figure}[htb]
\begin{center}
 \includegraphics[width=8cm]{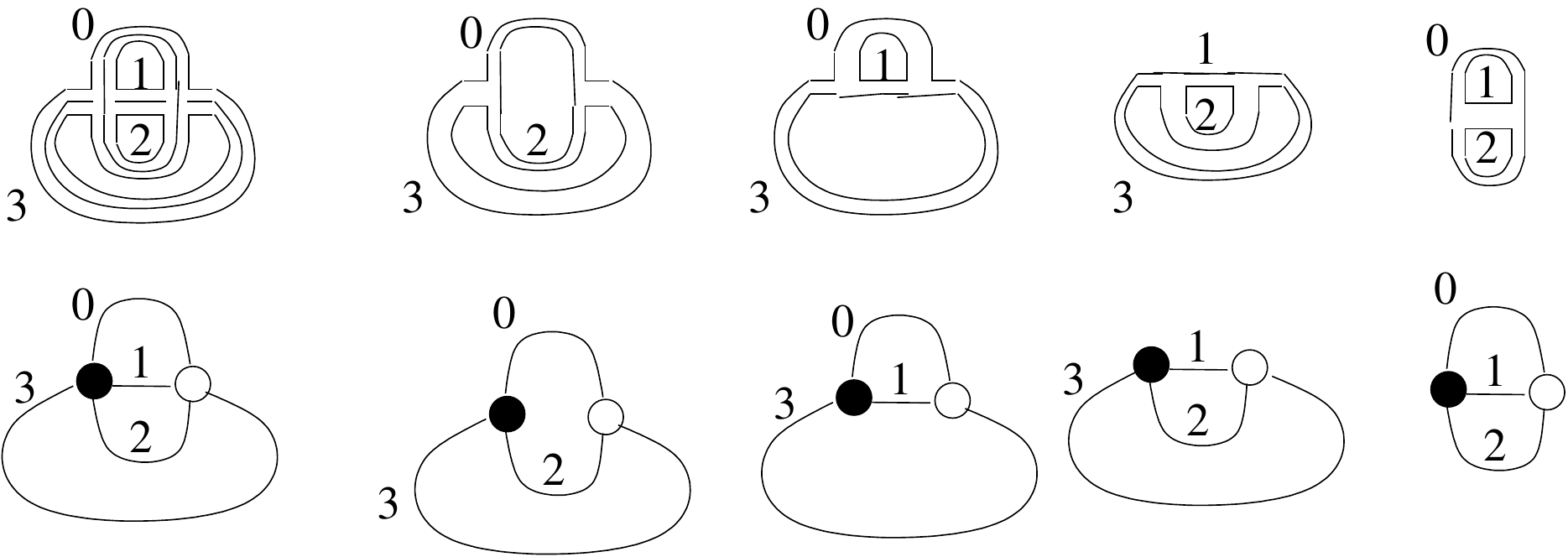}  
\caption{Bubbles of a graph in $D=3$.}
\label{fig:exemplubub}
\end{center}
\end{figure}

We denote $\widehat{i}_1 \dots \widehat{i}_{n}= \{0,\dots,D\} \setminus \{ i_1 \dots i_{n} \}$. 
To a colored graph $\cG$ one associates two categories of subgraphs: its bubbles \cite{color, lost} and 
its jackets \cite{colorN, colorNleadD,sefu3}.
The $0$-bubbles of $\cG$ are its vertices and the $1$-bubbles are its lines.
For $n\ge 2$, the {\bf $n$-bubbles} with colors $\{i_1,\dots,i_n\}$ of $\cG$
are the connected components (labeled $\rho$) obtained from $\cG$
by deleting the lines and faces containing at least one of the colors $
\widehat{i}_1 \dots \widehat{i}_{n}$. 
We denote $\cB^{i_1\dots i_n}_{(\rho)}$ the $n$-bubbles of colors $i_1\dots i_n$,
$\cB^{i_1\dots i_n}$ their number and $\cB^{[n]}=\sum_{i_1<i_2\dots<i_n} \cB^{i_1\dots i_n}$ 
the total number of bubbles with $n$-colors.
For $n\ge 2$ each $n$-bubble is a $n$-colored graph, and the $2$-bubbles are
the faces of $\cG$. An example of a $3+1$ CGFT graph in and its $3$-bubbles is given in figure
\ref{fig:exemplubub}. The $n$-bubbles of $\cG$ are the colored graphs dual to the links of the 
normal pseudo manifold dual to $\cG$ \cite{lost}.

\begin{figure}[htb]
\begin{center}
 \includegraphics[width=5cm]{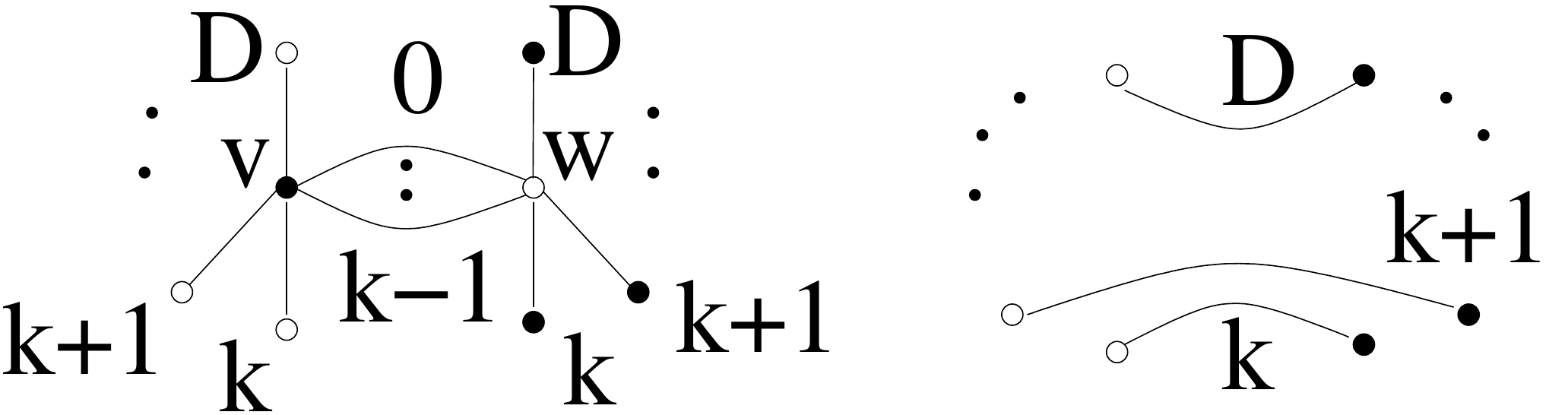}  
\caption{Contraction of a k-Dipole.}
\label{fig:1canc}
\end{center}
\end{figure}

Consider $k$ lines of colors $i_0,\dots, i_{k-1} $ connecting all the 
same two vertices $v$ and $w$ in a graph $\cG$. Each of the vertices $v$ and $w$ belongs to a 
$D+1-k$-bubble  $\cB^{\widehat{i}_0\dots \widehat{i}_{k-1}}_{(\alpha)} $ 
and $\cB^{\widehat{i}_0\dots \widehat{i}_{k-1} }_{ (\beta) }$. If the
two bubbles are {\it different} the lines form a {\bf $k$-Dipole} (denoted $d_k$)
\cite{colorN,FG,Lins}. 
A $k$-Dipole can be contracted, that is the lines 
$i_0\dots i_{k-1}$ together with the vertices $v$ and $w$ can be deleted from $\cG$  
and the remaining lines reconnected {\it respecting the coloring} 
(see figure  \ref{fig:1canc}). The inverse of the $k$-Dipole contraction is called 
a $k$-Dipole creation\footnote{The contraction/creation of a $D$-Dipole is superfluous 
and can always be traded for the contraction/creation of a $1$-Dipole.}.
We denote $\cG/d_k$ the graph obtained from $\cG$ by contracting $d_k$.
We say that $\cG$ and $\cG/d_k$ are {\it combinatorially equivalent}, $\cG \sim^{(c)} \cG/d_k$.

Consider $d_1$ a $1$-Dipole of color $i$. The graph $\cG/d_1$ has two less vertices and one less $D$-bubble 
of colors $\widehat{i}$ than $\cG$. The number of 
$D$-bubbles of colors $\widehat{j}, \; j\neq i$ and the connectivity of $\cG$ are unchanged 
under this contraction. Contracting the maximal number of $1$-Dipoles $\cG$ 
reduces to some graph $\cG_f$ with $2p_f \ge 2$ vertices and one $D$-bubble 
for each $i$, $\cB_f^{[D]}=D+1$. We have
\bea\label{eq:smeche}
p- p_f = \cB^{[D]} - \cB^{[D]}_f \Rightarrow  p_f-1 = p + D - \cB^{[D]} \ge 0 \; .
\eea

The bubbles encode the cellular structure of the graph $\cG$ \cite{color,PolyColor,lost},
and their nested structure encodes its homology (and that of its dual pseudo manifold).
However, as the bubbles are themselves colored graphs, they are relatively difficult to handle. 

\begin{definition}
A colored {\bf jacket} $\cJ$ of $\cG$ is the ribbon graph made by the faces of colors 
$(\tau^q(0),\tau^{q+1}(0))$, for some cycle $\tau$ of $D+1$ elements,
modulo the orientation of the cycle. We denote $J=\{ (\tau^q(0), \tau^{q+1}(0)) \;  | k=0,\dots, D \}$
the set of faces of $\cJ$.
\end{definition}

\begin{figure}[htb]
\begin{center}
 \includegraphics[width=6cm]{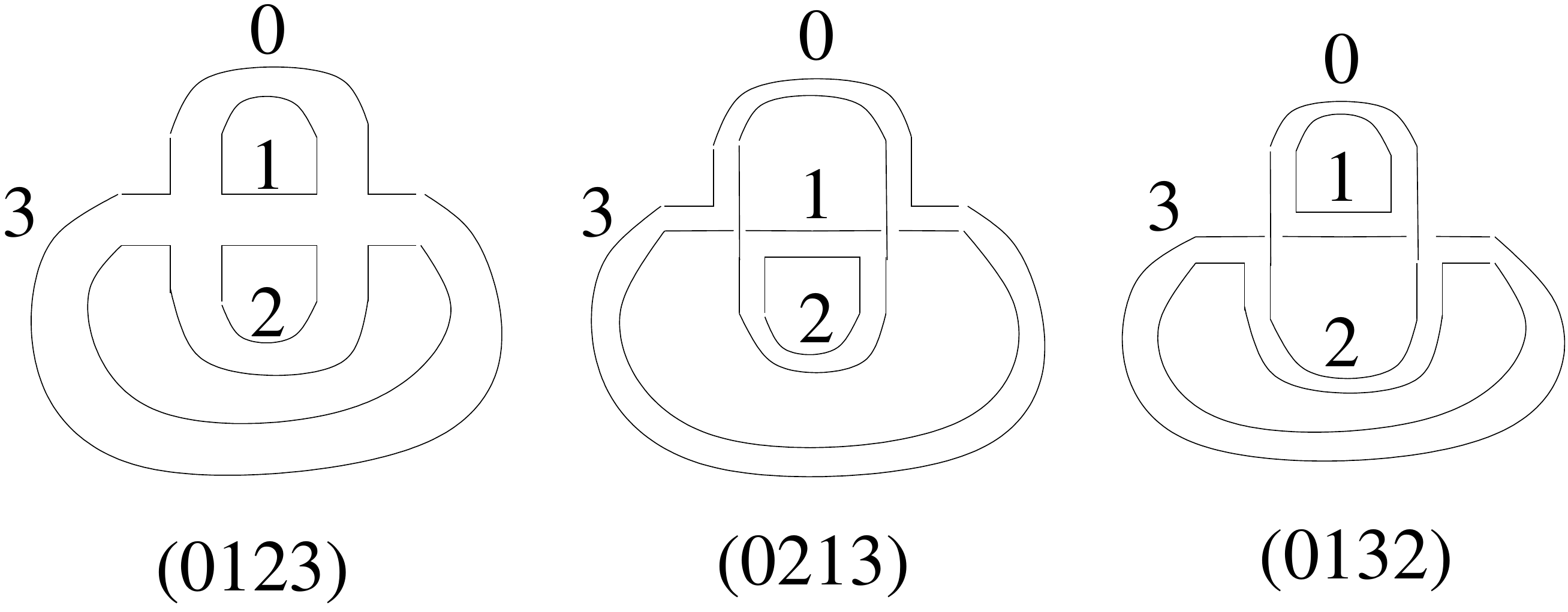}  
\caption{Jackets of a graph in $D=3$.}
\label{fig:exemplujacket1}
\end{center}
\end{figure}

An example of the jackets of a $3+1$ graph (and their associated cycles) is given in figure \ref{fig:exemplujacket1}.

The (unique) jacket of a $2+1$ colored graph is the graph itself.
The reader can check that $\cJ$ and $\cG$ have the same connectivity, the number of jackets
of a $D+1$ colored graph is $ \frac{1}{2}D!$ and the number of jackets containing a given face
is $(D-1)!$. The ribbon lines of the jacket $\cJ$ separate two faces, $(\tau^{-1}(j),j)$ and $(j,\tau(j))$ 
and inherit the color $j$ of the line in $\cG$. As the $n$-bubbles 
are $n$-colored graphs they also posses jackets which can be obtained from 
the jackets of $\cG$. 

For a jacket $\cJ$, $J^{\widehat{i}} = J \setminus \{(\tau^{-1}(i), i), (i,\tau(i))\} \cup \{ (\tau^{-1}(i), \tau(i)) \} $ 
is a cycle over the $D$ elements $\widehat{i}$. The ribbon subgraph of $\cG$ made of the faces in
$J^{\widehat{i} }$ is the union of several connected components, $\cJ^{\widehat{i}}_{(\rho)}$. Each
$\cJ^{\widehat{i}}_{(\rho)}$ is a jacket of the $D$-bubble 
$\cB^{ \widehat{i} }_{ (\rho) }$. Conversely, every jacket of $\cB^{ \widehat{i} }_{ (\rho) }$ is obtained 
from exactly $D$ jackets of $\cG$ corresponding to inserting the color $i$ anywhere along the cycle of $D$ elements.
We denote $g_{\cJ}$ the genus of the jacket $\cJ$.  For the examples of figures \ref{fig:exemplujacket1} and
\ref{fig:exemplubub}, the jacket $(0123)$ leads to the three jackets $(123)$, $(023)$ and $(012)$ 
one for each 3-bubble. Each of these jackets is in fact a bubble of figure \ref{fig:exemplubub},
being a $2+1$ colored graph.

\begin{definition}We define  
    \begin{itemize}
    \item The {\bf ``convergence degree''} (or simply {\bf degree}) of a graph $\cG$ is $\omega(\cG)=\sum_{\cJ} g_{\cJ}$,
    where the sum runs over all the jackets $\cJ$ of $\cG$. 
    \item The {\bf ``degree''} of a $k$-Dipole $d_k$ separating the bubbles $\cB^{\widehat{i}_0 \dots \widehat{i}_k}_{(\alpha)}$
   and $\cB^{\widehat{i}_0 \dots \widehat{i}_k}_{(\beta)}$ is $\omega (d_k) =
    \inf \Big{[} \omega(\cB^{\widehat{i}_0 \dots \widehat{i}_k}_{(\alpha)}),
    \omega(\cB^{\widehat{i}_0 \dots \widehat{i}_k}_{(\beta)}) \Big{]} $.
   \end{itemize}
\end{definition}

Having defined the required prerequisites, we will prove in the rest of this section a number of lemmas concerning CGFT graphs
which are the foundation of the $1/N$ expansion. First, the degree of a graph and of its bubbles are not independent.

\begin{lemma}\label{lem:degjackets} 
   The degrees of a graph $\omega(\cG)$ and its $D$-bubbles $\omega(\cB^{\widehat{i}}_{(\rho)})$ respect
   \bea
    \omega(\cG) = \frac{(D-1)!}{2} \Big{(} p+D-\cB^{[D]} \Big{)} + \sum_{i;\rho} \omega(\cB^{\widehat{i}}_{(\rho)}) \; .
   \eea
    In particular, by eq. \eqref{eq:smeche}, $p+D-\cB^{[D]} \ge 0$, thus
     $\omega(\cG)=0 \Rightarrow \omega(\cB^{\widehat{i}}_{(\rho)}) = 0 \; \forall i,\rho$ and $ p+D-\cB^{[D]} =0$.
\end{lemma}

\noindent{\bf Proof:}
The number of lines of $\cG$ and of any of its jackets is $\cL_{\cG}=(D+1)p$, hence the number of faces of the 
jacket $\cJ$ is $\cF_{\cJ} = (D-1) p + 2-2g_{\cJ}$.
Taking into account that $\cG$ has $\frac{1}{2} D!$ jackets and each face belongs to $(D-1)!$ jackets, 
\bea\label{eq:faces}
 \cF_{\cG} = \frac{D (D-1)}{2} p + D - \frac{2}{(D-1)!} \sum_{\cJ} g_{\cJ} = 
\frac{D (D-1)}{2} p + D - \frac{2}{(D-1)!} \omega(\cG) \; .
\eea
Each of the $D$-bubbles $\cB^{\widehat{i}}_{(\rho)}$ (with $\cN_{ \cB^{\widehat{i}}_{(\rho)} } = 2 p^{ \widehat{i}}_{(\rho)}$)
is a $D$-colored graph thus eq. \eqref{eq:faces} also holds for each $D$-bubble 
\bea\label{eq:faces2}
 \cF_{ \cB^{\widehat{i}}_{(\rho)} } = \frac{(D-1)(D-2)}{2} p^{ \widehat{i}}_{(\rho)} 
+ (D-1) - \frac{2}{(D-2)!} \omega(\cB^{\widehat{i}}_{(\rho)}) \; .
\eea
Each vertex of $\cG$ contributes to $D+1$ of its $D$-bubbles ($\sum_{i;\rho} p^{\widehat{i}}_{\rho} = (D+1)p$), 
and each face to $D-1$ of them. Adding eq. \eqref{eq:faces2} and dividing by $D-1$ yields
\bea
 \cF_{\cG} = \frac{(D-2)(D+1)}{2} p + \cB^{[D]} - \frac{2}{(D-1)!}  \sum_{i;\rho}  \omega(\cB^{\widehat{i}}_{(\rho)}) \; ,
\eea
which equated with \eqref{eq:faces} proves the lemma.

\qed

Second, the degree of a graph is preserved under $1$-Dipole contractions. More generally, we have

\begin{lemma}\label{lem:degreecontr}
   The degree of $\cG$ and $\cG/d_k$ are related by 
   \bea
    \omega(\cG) = \frac{(D-1)!} {2} \Big{(} (D+1)k - k^2 - D \Big{)} + \omega(\cG/d_k) \; .
   \eea    
 \end{lemma}

\noindent{\bf Proof:} 
Denote $\cB^{\widehat{i}_0\dots \widehat{i}_{k-1}}_{(\alpha)} $ 
and $\cB^{\widehat{i}_0\dots \widehat{i}_{k-1} }_{ (\beta) }$ 
the two $D$-bubbles separated by $d_k$. Under the contraction all 
faces $(p,q) \; p,q =i_0,\dots i_{k-1}$ are deleted
and all faces $(p,q) \; p,q \neq  i_0,\dots i_{k-1}$
are merged two by two, thus
\bea
  \cF_{\cG/d_k} = \cF_{\cG} - \frac{k(k-1)}{2} - \frac{(D-k+1)(D-k)}{2} \; .
\eea
But from eq. \eqref{eq:faces} we have
\bea
 &&\cF_{\cG} = \frac{D (D-1)}{2} p + D - \frac{2}{(D-1)!} \omega(\cG) \crcr
 &&\cF_{\cG/d_k} = \frac{D (D-1)}{2} (p-1) + D - \frac{2}{(D-1)!} \omega(\cG/d_k) \; ,
\eea
hence
\bea
\frac{2}{(D-1)!}\omega(\cG)  &=& 
\frac{2}{(D-1)!}\omega(\cG/d_k)  +  \frac{D (D-1)}{2} - \frac{k(k-1)}{2} - \frac{(D-k+1)(D-k)}{2} \crcr
&=& \frac{2}{(D-1)!}\omega(\cG/d_k)  - D - k^2+ k + Dk  \; .
\eea

\qed

One can track in detail the effect of a $1$-Dipole contraction over the degrees of the D-bubbles of a graph.
Denote $d_1$ a $1$-Dipole of color $i$ separating the two $D$-bubbles $\cB^{\widehat{i}}_{(\alpha)}$ and 
$\cB^{\widehat{i}}_{(\beta)}$. For $i\neq j$ or $i=j, \; \rho\neq \alpha,\beta$ we denote 
$\cB^{\widehat{j}}_{(\rho)}$ (resp. $\cB^{\widehat{j}}_{(\rho)}/d_1$) the $D$-bubbles
before (resp. after) contraction of $d_1$. The two $D$-bubbles $\cB^{\widehat{i}}_{(\alpha)}$ and 
$\cB^{\widehat{i}}_{(\beta)}$ are merged into a unique bubble $\cB^{\widehat{i}}_{(\alpha)\cup(\beta)}/d_1$
after contraction.

\begin{lemma}\label{lem:degbub}
   The degrees of the bubbles before/after contraction of a $1$-Dipole respect
   \bea
    && \omega( \cB^{\widehat{j}}_{(\rho)}/d_1) = \omega( \cB^{\widehat{j}}_{(\rho)}) \; ,  \qquad  i\neq j \; 
        \text{or}  \;
    i=j, \; \rho\neq \alpha,\beta \crcr
    && \omega( \cB^{\widehat{i}}_{(\alpha)\cup(\beta)}/d_1) = \omega(\cB^{\widehat{i}}_{(\alpha)} ) + 
     \omega( \cB^{\widehat{i}}_{(\beta)} ) \; .
   \eea
\end{lemma}

\noindent{\bf Proof:}
   Only the  $D$-bubbles containing one (or both) of the end vertices $v$ and $w$ of $d_1$ are affected by 
   the contraction.

   For $j\neq i$, any jacket $\cJ^{\widehat{j}}_{(\rho)}/d_1$ of $\cB^{\widehat{j}}_{(\rho)}/d_1$ has
    $2$ vertices less, $D+1$ lines less and $D-1$ faces less  
   than the corresponding jacket $\cJ^{\widehat{j}}_{(\rho)}$ of $\cB^{\widehat{j}}_{(\rho)}$. Therefore 
   $g_{\cJ^{\widehat{j}}_{(\rho)}/d_1}= g_{\cJ^{\widehat{j}}_{(\rho)}}$ and the degree is conserved.

   Any jacket $\cJ^{\widehat{i}}_{(\alpha)\cup(\beta)}/d_1 $ of the $D$-bubble $\cB^{\widehat{i}}_{(\alpha)\cup(\beta)}/d_1$, is
obtained by gluing two jackets $ \cJ^{\widehat{i}}_{(\alpha)}$ and $\cJ^{\widehat{i}}_{(\beta)}$ of 
$ \cB^{\widehat{i}}_{(\alpha)}$ and $\cB^{\widehat{i}}_{(\beta)}$. It follows  $\cJ^{\widehat{i}}_{(\alpha)\cup(\beta)}/d_1 $
has $2$ vertices  less, $D$ lines  less, $D$ faces  less and $1$ connected component  less than the two jackets
$ \cJ^{\widehat{i}}_{(\alpha)}$ and $\cJ^{\widehat{i}}_{(\beta)}$. Hence 
$g_{\cJ^{\widehat{i}}_{(\alpha)\cup(\beta)}/d_1 }=g_{\cJ^{\widehat{i}}_{(\alpha)}} +g_{\cJ^{\widehat{i}}_{(\beta)}}$, and the 
lemma follows.

\qed

These lemmas we proved so far are sufficient to perform the combinatorial $1/N$ expansion. We now include the topology in 
the picture. The fundamental combinatorial topology result we will use in the sequel \cite{FG,Lins} is that the 
two pseudo manifolds dual to $\cG$ and $\cG/d_k$ are homeomorphic if one of the bubbles 
$\cB^{\widehat{i}_0\dots \widehat{i}_{k-1}}_{(\alpha)} $ 
or $\cB^{\widehat{i}_0\dots \widehat{i}_{k-1} }_{ (\beta) }$ is dual to a {\it sphere} $S^{D-k}$.
We say that two such graphs are {\it topologically equivalent}, $\cG \sim^{(t)} \cG/d_k$ (and that the $k$-Dipole $d_k$ 
separates a sphere). As mentioned in section \ref{sec:mm}, it is in principle very difficult to check whether 
a graph is a sphere or not. However we can establish the following partial result 

\begin{lemma} \label{lem:sph}
 If $\omega(\cG)=0$ then $\cG$ is dual to a sphere $S^D$. The reciprocal holds in $D=2$.
\end{lemma}

\noindent{\bf Proof:} We use induction on $D$. In $D=2$, $\cG$ is a ribbon graph and
its degree equals its genus. For $D\ge 3$, as $\omega(\cG)=0$, lemma \ref{lem:degjackets} 
implies $\omega(\cB^{\widehat{i}}_{(\rho)})=0$ and, by the induction hypothesis,
all the bubbles $\cB^{\widehat{i}}_{(\rho)} $ are dual to spheres $S^{D-1}$. 
Any $1$-Dipole will separate a sphere, $\cG/d_1 \sim^{(t)} \cG$ and by lemma 
\ref{lem:degreecontr}, $\omega(\cG/d_1)=\omega(\cG)=0$. 
We iteratively contract a full set of $1$-Dipoles to reduce 
$\cG$ to a final graph $\cG_f \sim^{(t)}\cG$, $\omega(\cG_f)=0$ such that $\cG_f$ does not posses any 1-Dipoles.
It follows that $\cG_f$ has $\cB^{[D]}_f=D+1$ remaining $D$-bubbles (one for each colors $\widehat{i}$)
and, by lemma \ref{lem:degjackets}, $p_f+D-\cB_f^{[D]}=0$. We conclude that $p_f=1$ and $\cG_f$ represents the 
coherent identification of two $D+1$ simplices along their boundary i.e. it is dual to 
a sphere $S^D$.

\qed 

\section{Colored GFT Models}\label{sec:models}

Let $G$ be some compact multiplicative Lie group, and denote $h$ its elements,
$e$ its unit, and $\int dh$ the integral with respect to the Haar measure.
We denote $\delta^N(e)$ the delta function with some cutoff $N$ evaluated at the identity.
The large parameter $N$ can be either a cutoff on the discrete Fourier modes,
or a cutoff in a heat kernel regularization \cite{Geloun:2011cy,BonzSmer}. Irrespective of its particular implementation,
$\delta^N(e)$ diverges when $N \to \infty$. 

Let $\bar \psi^i,\psi^i$,  $i=0,1,\dots, D$ be $D+1$ couples of complex 
scalar (or Grassmann) fields over $D$ copies of $G$, $\psi^i:G^{\times D} \rightarrow \mathbb{C}$. 
The index $i\in \{0,\dots,D\}$ of each field is the {\it color} index.
We denote $ \psi(h_0,\dots, h_{D-1}):=\psi_{h_0,\dots, h_{D-1}}$.

The partition function of the $D$ dimensional colored GFT \cite{color,lost,colorN} is defined by the path integral 
\bea\label{eq:partition}
 e^{-F(\lambda,\bar\lambda)} = Z(\lambda,\bar\lambda) = \int \prod_{i=0}^D d\mu_P(\psi^i,\bar\psi^i) \; e^{-S^{int}-\bar S^{int}}
\; ,
\eea
with normalized Gaussian measure of covariance $P$ and interaction $S^{int}$.
The free energy $F$ of the CGFT writes as a sum over connected vacuum 
colored graphs $\cG$. A model is a choice of the covariance $P$
and of the interaction $S^{int}$.

\section{The Independent Identically Distributed Model}\label{sec:iid}

The simplest model is the independent identically distributed  (i.i.d.) model defined by
\bea\label{eq:actioniid}
&& P_{h_{0}\dots h_{D-1} ; h_{0}'\dots h_{D-1}'}  = \prod_{i=0}^{D-1} \delta^N \bigl( h_{i} (h_{i}')^{-1} \bigr)  \;, \crcr
&& S^{int} = \frac{\lambda}{ \sqrt{ \delta^N(e)^{\frac{D(D-1)}{2}} } } 
\int \prod_{i<j} dh_{ij} \; \prod_{i=0}^{D+1} \psi^i_{h^{}_{ii-1} \dots h^{}_{i0} h^{}_{iD} \dots h_{ii+1}}  \; .
\eea
where $h_{ij} = h_{ji}$. For the i.i.d model with $G=U(1)$ one can rewrite eq. \eqref{eq:partition}
in Fourier modes and recover just the straightforward generalization of a i.i.d. matrix model to tensors with
$D$ indices.

\begin{figure}[htb]
\begin{center}
 \includegraphics[width=2cm]{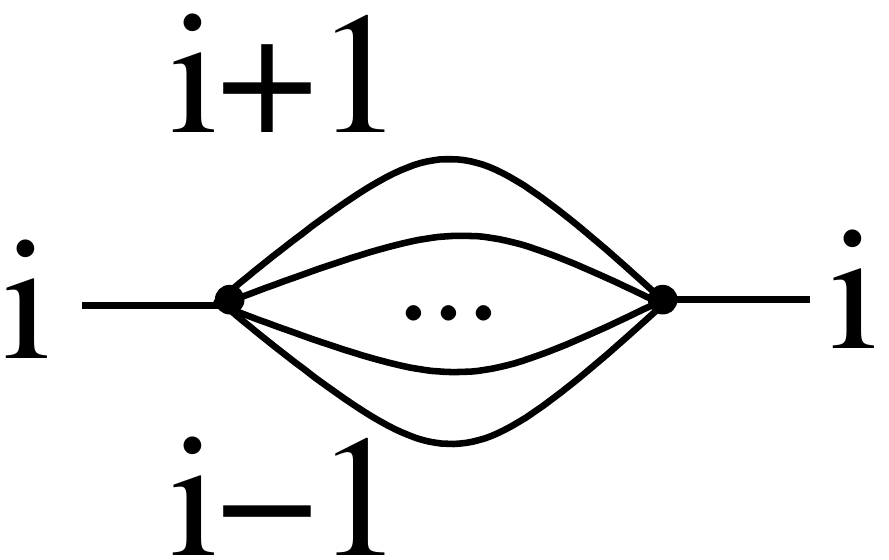}  
\caption{The two point colored graph $\cS$.}
\label{fig:melon}
\end{center}
\end{figure}

Consider the two point colored graph with two vertices connected by $D$ lines (denoted $\cS$)
represented in figure \ref{fig:melon}. For any graph $\cG$, one can consider
the family obtained by inserting $\cS$ an arbitrary number of times on 
any line of $\cG$ . The scaling of the coupling constants in eq. \eqref{eq:actioniid} is the only scaling 
which ensures that this family has uniform degree of divergence, hence it is the only scaling 
under which a $1/N$ expansion makes sense.

\subsection{Amplitude} 

For an arbitrary connected graph $\cG$ the $\delta^N$ functions compose along the faces.
From now on we drop the superscript $N$ on the $\delta$ functions, and recall that our large parameter is 
$\delta(e)$.
The amplitude of a CGFT graph of the i.i.d. model is
\bea\label{eq:ampliiid}
A^{\text{i.i.d.}}(\cG) = (\lambda \bar\lambda)^{p} \; \delta(e)^{ - p \frac{D(D-1)}{2} + \cF_{\cG}} = 
(\lambda \bar\lambda)^{p} \; \delta(e)^{D - \frac{2}{(D-1)! } \omega(\cG) }\; .
\eea
where we used equation \eqref{eq:faces}. For the i.i.d. model, the convergence degree,
$\omega(\cG)$, plays in higher dimensions the same role the genus played for matrix models.
The $1/N$ expansion of the free energy of the i.i.d. model is an expansion in the degree 
\bea
 F^{\text{i.i.d.}}(\lambda,\bar\lambda) = \sum_{\omega=0}^{\infty} C^{[\omega]}(\lambda ,\bar\lambda)
\; \delta(e)^{D - \frac{2}{(D-1)!}\omega} \qquad
C^{[\omega]}(\lambda ,\bar\lambda) = 
\sum_{\cG, \; \omega(\cG) = \omega } \frac{ 1 } {s(\cG)} (\lambda \bar\lambda)^{\cN_{\cG}/2}  \; ,
\eea 
with $s(\cG)$ some symmetry factor. The coefficient $C^{[\omega]}(\lambda, \bar\lambda)$ is the sum of 
the series of graphs of degree $\omega$.
The fundamental difference between the two dimensional case and the general case is that, whereas
the genus is a topological invariant, the degree {\bf is not}. This leads to the two distinct expansions,
 the ``combinatorial $1/N$ expansion'' and the ``topological $1/N$ expansion'' mentioned in the introduction. 
Note however that the leading order is given by graphs of degree $0$, which are, from lemma \ref{lem:sph}, 
topological spheres \cite{colorNleadD}.

\subsection{The Combinatorial $1/N$ Expansion}

From lemma \ref{lem:degreecontr} the degree of a graph is {\it invariant} under arbitrary
$1$-Dipole contractions. This can be used to give an alternative characterization of 
$C^{[\omega]}(\lambda, \bar\lambda)$, which can be more useful when studying the critical 
behavior of the model. For any graph $\cG$ we reduce a maximal number of $1$-Dipoles
to obtain a simpler graph with the same amplitude as $\cG$.

\bigskip

\noindent{\bf Combinatorial Bubble Routing.} 
For each color $i$ we designate one of the $D$-bubbles $\cB^{\widehat{i}}_{(\rho)} $
as root $\cR^{\widehat{i}}_{(1)}$. The total number of roots of a graph is $\cR^{[D]} = D+1$.
We associate to the bubbles $\widehat{i}$ of $\cG$ 
a ``$\widehat{i}$ connectivity graph''. Its vertices represent
the various bubbles $\cB^{\widehat{i}}_{(\rho)} $. Its lines are the lines of color $i$ in $\cG$. 
They either start and end on the same bubble $ \cB^{\widehat{i}}_{(\alpha)} $ 
(``tadpole'' lines in the connectivity graph), or not. 
A particularly simple way to picture the $\widehat{i}$ connectivity
graph is to draw $\cG$ with the lines $j\neq i $ much shorter than the lines $i$.
We chose a tree in the connectivity graph, $\cT^{i}$ (and call the rest of the lines $i$ 
``loop lines''). All the $\cB^{\widehat{i}} - 1 $ lines of $\cT^i$ are 1-Dipoles and we contract them. 
We end up with a connectivity graph with only one vertex corresponding tho the root bubble $\cR^{\widehat{i}}_{(1)}$. 
The remaining lines of color $i$ cannot be contracted further (they are tadpole lines in the 
connectivity graph). The number of the $D$-bubbles of the other colors is unchanged 
under these contractions.

We iterate for all colors starting with $D$. The routing tree $\cT^{j}$ is chosen in the graph 
obtained {\it after} contracting $\cT^{j+1}, \dots \cT^D $. 
The number of bubbles of colors $q>j$ are constant under contractions of
$1$-Dipoles of color $j$, hence the latter {\it cannot} create new 1-Dipoles of color $q$. 
Reducing a full set of 1-Dipoles indexed by $D+1$ routing trees $\cT^0, \dots \cT^{D+1} $ 
we obtain a graph whose all bubbles are roots, called a 
{\it combinatorial core graph}.

\begin{definition}\label{def:corecomb}{\bf [Combinatorial Core Graph]}
   A $D+1$ colored graph $\cG^{(c)}_p$ with $2p$ vertices is called a combinatorial core graph at order $p$ if, 
   for all colors $i$, it has a unique $D$-bubble $\cR^{\widehat{i}}_{(1)}$.
\end{definition}

Note that in fact we already used the combinatorial core graphs twice: first to prove eq. \ref{eq:smeche} and second
to prove lemma \ref{lem:sph}. The important feature of combinatorial core graphs is that their degree (and consequently amplitude)
admits a bound in the number of vertices $p$: by lemma \ref{lem:degjackets} 
    \bea
    \omega(\cG^{(c)}_p) = \frac{(D-1)!}{2} \Big{(} p+D-\cR^{[D]} \Big{)} + \sum_{i;\rho} \omega(\cR^{\widehat{i}}_{(\rho)})         
    \ge \frac{(D-1)!}{2} (p-1)\; .
   \eea

\bigskip

\noindent{\bf Combinatorial core equivalence classes}.
The combinatorial core graph one obtains by routing is {\it not} independent of the routing trees. 
The same graph leads to several {\it equivalent} combinatorial core graphs, all {\it at the same
order} $p$, and {\it with the same amplitude} 
:$\cG^{(c)}_p \sim^{(c)} \cG^{(c)}_p{}'$ and 
$A^{\text{i.i.d.}} (\cG^{(c)}_p)= A^{\text{i.i.d.}} (\cG^{(c)}_p{}')$. We call two such core graphs
{\it combinatorial core equivalent}, $\cG^{(c)}_p \simeq^{(c)} \cG^{(c)}_p{}'$.
The amplitudes of $\cG$ and $\cG^{(c)}_p$ are related by
\bea\label{eq:famcomb}
 A^{\text{i.i.d.}}(\cG)= (\lambda \bar\lambda)^{ \cB^{[D] } - D-1 } A^{\text{i.i.d.}} (\cG^{(c)}_{p} )  \;, 
  \qquad p= \frac{\cN_{\cG}}{2} - \bigl(\cB^{ [D]  }- D-1  \bigr) \; .
\eea

An arbitrary graph will route to a unique combinatorial core equivalence class. The $1/N$ expansion
of the free energy of the colored GFT in arbitrary dimensions can be recast in combinatorial 
core equivalence classes 
\bea\label{eq:free}
 F^{\text{i.i.d.}}(\lambda, \bar\lambda) = \sum_{p=1}^{\infty} 
\sum_{  \omega= \frac{(D-1)!}{2} ( p-1 ) } \Big{[}\sum_{[\cG^{(c)}_p] \; , \omega(\cG^{(c)}_p)=\omega}
 C^{[\cG^{(c)}_p]}(\lambda,\bar\lambda) \Big{]}  \; \delta(e)^{D - \frac{2}{ (D-1)! } \omega } \; ,
\eea
where $C^{[\cG^{(c)}_p]}( \lambda,\bar\lambda )$ counts all the graphs
routing (via a combinatorial bubble routing) to the equivalence class $[\cG^{(c)}_p]$. The crucial feature of the
expansion in combinatorial core classes is that it can be evaluated order by order. 
One needs to draw all core graphs at order $p$, compute their amplitudes, and divide them into combinatorial
equivalence classes. The last step, identifying the combinatorial equivalence classes, is potentially 
difficult, but one needs to deal with this problem only a {\it finite} number of time in order to 
write all the terms up to a given order.

\subsection{Topological $1/N$ expansion}\label{sec:topo1/N}

To write a series taking into account not only the amplitude but also the topology of the graphs one must rely 
on topological equivalence rather then combinatorial equivalence.
This means that, instead of contracting all $1$-Dipoles, one should contract only $1$-Dipoles separating spheres. Of 
course it is out of the question to reduce all such $1$-Dipoles for an arbitrary graph (one would need to deal with 
a very difficult problem for every graph $\cG$). We will modestly only reduce the $1$-Dipoles of degree $0$. Not 
only such contractions preserve the topology (from lemma \ref{lem:sph}), but also they 
lead to a well defined $1/N$ expansion. Unsurprisingly the final series will have some degree of redundancy
which we will detail later.

\bigskip

\begin{figure}[htb]
\begin{center}
 \includegraphics[width=4cm]{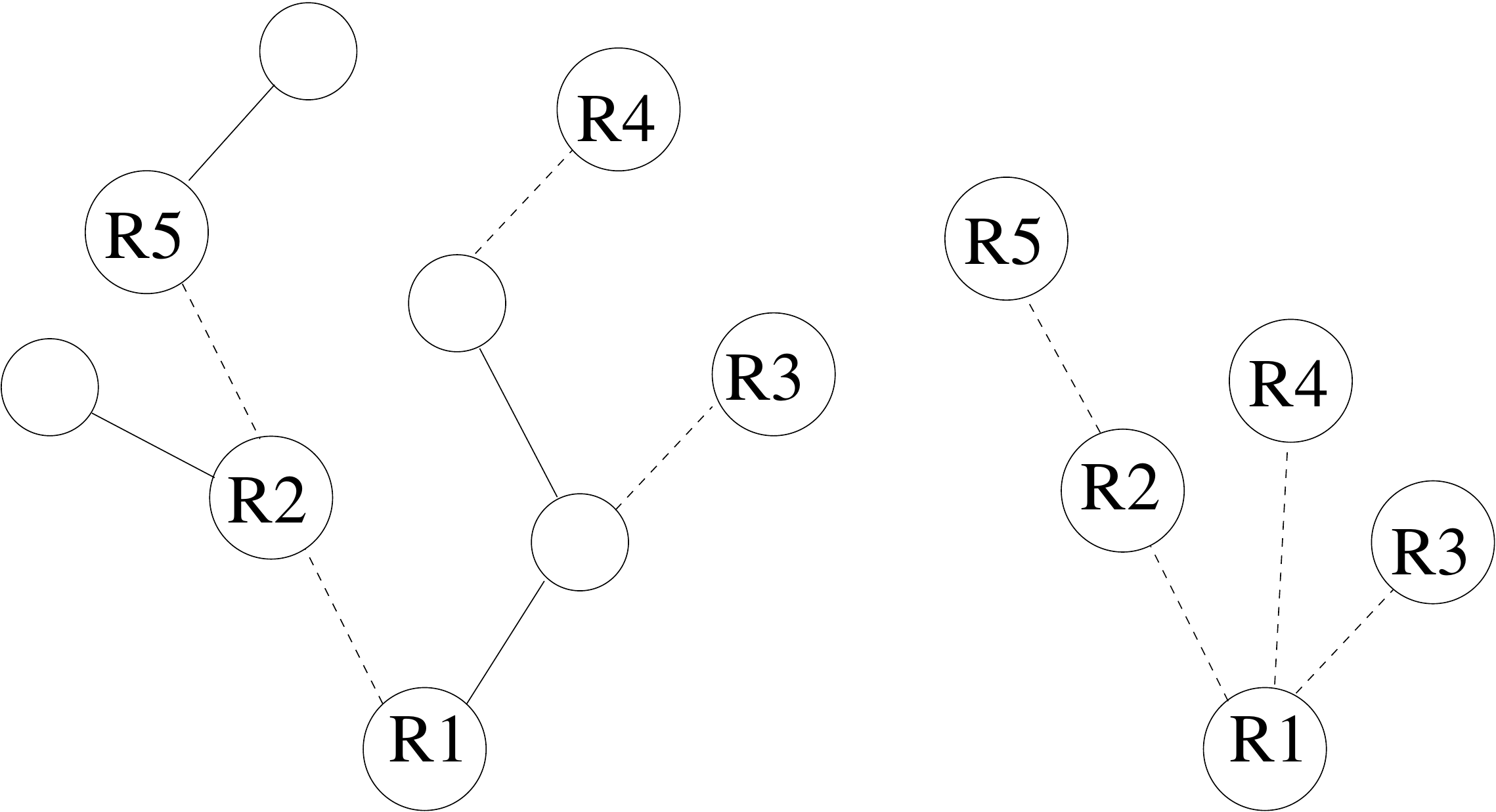}
\caption{A tree $\cT^3$ in the $012$ connectivity graph.}
\label{fig:treebub}
\end{center}
\end{figure}

\noindent{\bf Topological Bubble Routing.} 
We start by choosing a set of roots of $\cG$. Consider a color $i$. 
If there exist $D$-bubbles of degree $\omega(\cB^{\widehat{i}}_{(\rho)}) \ge 1 $
we set all of them as roots $\cR^{\widehat{i}}_{(\mu)} $. If not (i.e. if $ \omega(\cB^{\widehat{i}}_{(\rho)}) =0$
for all $\rho$) we choose one of them as root $\cR^{\widehat{i}}_{(1)}$. 
We call $\cR^{\widehat{i}}_{(1)} $ the principal root and $\cR^{\widehat{i}}_{(\mu)}, \; \mu >1 $
``branch roots''. Iterating for all colors we identify all the roots of $\cG$. 
We denote $\cR^{\widehat{i}}$ the number of roots of color $\widehat{i}$ and 
$\cR^{[D]} = \sum_i \cR^{\widehat{i}}$ .
We chose a tree in the connectivity graph $\widehat{i}$, $\cT^{i}$. For a branch root 
$ \cR^{\widehat{i}}_{(\mu)},\; \mu>1$, 
the line incident on it and belonging to the unique path in 
$\cT^{i}$ connecting $\cR^{\widehat{i}}_{(\mu)} $ to the principal root 
$\cR^{\widehat{i}}_{(1)} $ is represented as dashed.
All the other lines in $\cT^{i}$ are represented as solid lines. An example is given in 
figure \ref{fig:treebub}.

All the $\cB^{\widehat{i}} - \cR^{\widehat{i}} $ solid lines in $\cT^i$ are 1-Dipoles of degree $0$ and we contract them. 
We end up with a connectivity graph with vertices corresponding to the roots 
$\cR^{\widehat{i}}_{(\mu)} $. The remaining 
lines of color $i$ cannot be contracted further (they are either tadpole lines or they 
separate two roots, hence are not 1-Dipoles of degree 0). 
Neither the number nor the topology of the bubbles of the other colors is changed under these contractions.

We iterate for all colors starting with $D$. 
The routing tree $\cT^{j}$ is chosen in the graph obtained {\it after} contracting $\cT^{j+1}, \dots \cT^D $. 
The degrees $\omega(\cR^{\widehat{q}}_{(\rho)}), \; q>j$ are unaffected by the contractions of
$1$-Dipoles of color $j$ (by lemma \ref{lem:degbub}), hence the latter {\it cannot} create new 1-Dipoles of degree 0 
and color $q$. Reducing a full set of 1-Dipoles of degree $0$ indexed by $D+1$ routing trees
$\cT^0, \dots \cT^{D+1} $ we obtain a graph whose all bubbles are roots, called a 
{\it topological core graph}.

\begin{definition}\label{def:coretop}{\bf [Topological Core Graph]}
   A $D+1$ colored graph $\cG^{(t)}_p$ with $2p$ vertices is called a topological core graph at order $p$ if, for 
all colors $i$,
   \begin{itemize}
    \item either $\cG$ has a unique $D$-bubble $\cR^{\widehat{i}}_{(1)}$ of degree $\omega(\cR^{\widehat{i}}_{(1)})=0$.
    \item or all bubbles $\cR^{\widehat{i}}_{(\rho)}$, have degree $\omega(\cR^{\widehat{i}}_{(\rho)}) \ge 1$ . 
   \end{itemize}
\end{definition}

As for combinatorial core graphs, the degree (and amplitude) of topological core graphs admit a bound 
in the number of vertices $p$: by lemma \ref{lem:degjackets} 
    \bea
    \omega(\cG^{(t)}_p) = \frac{(D-1)!}{2} \Big{(} p+D-\cR^{[D]} \Big{)} + \sum_{i;\rho} \omega(\cR^{\widehat{i}}_{(\rho)}) \; ,
   \eea
   and the definition of a topological core graph implies 
$ \sum_{i;\rho} \omega(\cR^{\widehat{i}}_{(\rho)}) \ge \cR^{[D]}- (D+1)$, hence
   \bea\label{eq:degtopol}
   \omega(\cG^{(t)}_p) \ge  \Big{(}\frac{(D-1)!}{2} -1 \Big{)} \Big{(} p+D-\cR^{[D]} \Big{)} +p-1 \ge p-1 \; ,
   \eea 
   where for the last inequality we used again eq. \eqref{eq:smeche}.

\bigskip

\noindent{\bf Topological core equivalence classes}.
The core graph one obtains by routing is again not independent of the routing trees. 
The same graph leads to several equivalent core graphs, all {\it at the same
order} $p$, {\it topologically equivalent} and {\it with the same amplitude} 
($\cG^{(t)}_p \sim^{(t)} \cG^{(t)}_p{}'$ and $A^{\cG^{(t)}_p}=A^{\cG^{(t)}_p{}'}$). We call two such core graphs
{\it topological core equivalent}, $\cG^{(t)}_p \simeq^{(t)} \cG^{(t)}_p{}'$.
The amplitude of $\cG$ and $\cG^{(t)}_p$ are related by
\bea\label{eq:famtop}
 A^{\text{i.i.d.}} (\cG) = (\lambda \bar\lambda)^{ \cB^{[D] } - \cR^{[D]} } A^{\text{i.i.d.}} (\cG^{(t)}_{p} ) \;, 
  \qquad p= \frac{\cN_{\cG}}{2} - \bigl(\cB^{[D]  }- \cR^{[D]  } \bigr) \; .
\eea

An arbitrary graph will route to a unique topological core equivalence class. The $1/N$ expansion
of the free energy of the colored GFT in arbitrary dimensions admits an expansion in topological core equivalence 
classes 
\bea\label{eq:topoiid}
 F^{\text{i.i.d.}}(\lambda, \bar\lambda) = \sum_{p=1}^{\infty} 
\sum_{  \omega=  p-1  } \Big{[}\sum_{[\cG^{(t)}_p] , \; \omega(\cG^{(t)}_p)=\omega}
 C^{[\cG^{(t)}_p]}(\lambda,\bar\lambda) \Big{]}  \; \delta(e)^{D - \frac{2}{ (D-1)! } \omega } \; ,
\eea
where $C^{[\cG^{(t)}_p]}( \lambda,\bar\lambda )$ counts all the graphs
routing (via a topological bubble routing) to $[\cG^{(t)}_p]$. The classes $[\cG^{(t)}_p]$ can again be 
listed order by order. Again identifying the topological equivalence classes at order $p$ is 
difficult, but one needs to deal with this problem 
only a finite number of times.

\section{The Boulatov Ooguri Model}\label{sec:topomodel}

The second model we will detail in this paper is the Boulatov Ooguri (BO) topological model defined by 
\bea\label{eq:action}
&& P_{h_{0}\dots h_{D-1} ; h_{0}'\dots h_{D-1}'} 
 = \int dh \;  \prod_{i=0}^{D-1} \delta^{N}\bigl( h_{i} h (h_{i}')^{-1} \bigr) \; ,\crcr
&& S^{int} = \frac{\lambda}{ \sqrt{ \delta^N(e)^{\frac{(D-2)(D-1)}{2}} } } 
\int \prod_{i<j} dh_{ij} \; \prod_{i=0}^{D+1} \psi^i_{h^{}_{ii-1} \dots h^{}_{i0} h^{}_{iD} \dots h_{ii+1}}  \; , 
\eea

 The amplitude of $\cG$, is
\cite{color,colorN,sefu3}
\bea  \label{eq:ampli}
A^{\text{BO} } (\cG) = \frac {  (\lambda\bar\lambda)^p}
{ [\delta^N(e)]^{p\frac{(D-2)(D-1)}{2} } }  
\int \prod_{\ell\in \cL_{\cG}} dh_{\ell} 
\prod_{f\in \cF_{\cG}} \delta^N_{f}(\prod_{\ell\in f }^{\rightarrow} h_{\ell}^{\sigma^{ \ell | f}} )
\; ,
\eea
where $\sigma^{\ell|f}=1$ (resp. $-1$) if the orientations of 
$\ell$ and $f$ coincide (resp. are opposite) and $\sigma^{\ell|f}=0$ if $\ell$ does not 
belong to the face $f$. Unlike the i.i.d. model one cannot easily evaluate the amplitude 
for arbitrary graphs. However one can still build a topological expansion along the lines of the one
of the i.i.d. models.  Again we drop the superscript $N$ on $\delta^N$, and denote $=$ two 
quantities which are equal at leading order in $N$.

\subsection{Power counting}\label{sec:pcount}

In this section we establish a general power counting bound for the amplitude of an arbitrary graph
in the BO model in terms of its convergence degree $\omega(\cG)$. We start by a technical prerequisite.

\bigskip

\noindent{\bf Face routing of ribbon graphs.} 
The jackets (which are ribbon graphs) can be used to simplify the amplitude \eqref{eq:ampli} of 
a graph. To every ribbon graph $\cH$ one associates a dual graph $\tilde \cH$. 
The construction is standard (see for instance \cite{param} 
and references therein). The vertices of $\cH$, correspond to the faces of $\tilde \cH$, 
its lines to the lines of $\tilde \cH$ and its faces to the vertices of $\tilde \cH$.
The lines of $\cH$ admit (many) partitions in three disjoint sets: a tree $\cT$ in $\cH$, 
($|\cT|=|\cN_{\cH}|-1$), a tree $\tilde \cT$ in the its dual $\tilde \cH $, ($|\tilde \cT|=|\cF_{\cH}|-1$), 
and a set $\cL \setminus \cT \setminus \tilde \cT$, ($| \cL \setminus \cT \setminus \tilde \cT|=2g_{\cH}$ ) 
of ``genus'' lines (\cite{param}). 

We orient the faces of $\cH$ such that the two strands of every line have opposite orientations.
We set a face of $\cH$ as ``root'' (denoted $r$). Consider a face $f$ sharing some line 
$l(f,\tilde \cT) \in \tilde \cT$ with the root (that is the two strands of $l(f,\tilde \cT)$ belong one 
to $r$ and the other to $f$). The group element $h_{l(f,\tilde \cT)}$ appears exactly once in the argument of
$\delta_f$ and $\delta_r$ 
\bea\label{eq:rout1}
\delta_{r}(\prod_{\ell}^{\rightarrow} h_{\ell}^{\sigma^{\ell | r}}) \;
\delta_{f}(\prod_{\ell}^{\rightarrow} h_{\ell}^{\sigma^{\ell | f}})
=\delta_{r}\Big{(} (\prod_{\ell \neq l(f,\tilde \cT) }^{\rightarrow} h_{\ell}^{\sigma^{\ell | r}} ) 
h_{l(f,\tilde \cT) }^{\sigma^{ l(f,\tilde \cT) | r}} \Big{)} \;
\delta_{f} \Big{(} h_{ l (f,\tilde \cT) }^{\sigma^{l(f,\tilde \cT) | f}} 
(\prod_{\ell \neq l(f,\tilde \cT) }^{\rightarrow} h_{\ell}^{\sigma^{\ell | f}}) \Big{)} \; ,
\eea
where we set $l(f,\tilde \cT) $ as the last line of $r$ and as the first line of $f$. 
By our choice of orientations $\sigma^{l(f,\tilde \cT) | r }\sigma^{l(f,\tilde \cT) | f}=-1 $.
We change variables to  $\tilde h_{l(f,\tilde \cT)}^{\sigma^{l(f,\tilde \cT) | f} } = 
 h_{l(f,\tilde \cT)}^{\sigma^{l(f,\tilde \cT) | f}} 
(\prod_{\ell \neq l(f,\tilde \cT) }^{\rightarrow} h_{\ell}^{\sigma^{\ell | f}}) $, 
$d \tilde h_{l(f,\tilde \cT)} = d h_{l(f,\tilde \cT)} $, and
eq. (\ref{eq:rout1}) becomes
\bea
&& \delta_{r}\Big{(} (\prod_{\ell \neq l(f,\tilde \cT) }^{\rightarrow} h_{\ell}^{\sigma^{\ell | r}} )
(\prod_{\ell \neq  l(f,\tilde \cT) }^{\rightarrow} h_{\ell}^{\sigma^{\ell | f}} ) 
\;  \tilde h_{l(f,\tilde \cT)}^{\sigma^{l(f,\tilde \cT) | r} }  
\Big{)} \;
\delta_{f} \Big{(}  \tilde h_{l(f,\tilde \cT)}^{\sigma^{l(f,\tilde \cT) | f} }    \Big{)} 
\crcr
&&=\delta_{r}\Big{(} (\prod_{\ell \neq l(f,\tilde \cT) }^{\rightarrow} h_{\ell}^{\sigma^{\ell | r}} )
(\prod_{\ell \neq  l(f,\tilde \cT) }^{\rightarrow} h_{\ell}^{\sigma^{\ell | f}} ) 
\Big{)} \;
\delta_{f} \Big{(}  \tilde h_{l(f,\tilde \cT)}^{\sigma^{l(f,\tilde \cT) | f} }    \Big{)} \; .
\eea
This trivial change of variables has two consequences. First the face $f$ is canonically associated 
to the line $l(f,\tilde \cT)$. Second, the face $r$ becomes a root face in the graph 
$\cH-l(f,\tilde \cT)$, obtained from $\cH$ by deleting $l(f,\tilde \cT)$ (and connecting 
$r$ and $f$ into $r \cup f$, see figure \ref{fig:del}).
Iterating for all faces except the root we get
\bea \label{eq:routing}
 \prod_{f\in \cH} \delta_{f}( \prod_{\ell}^{\rightarrow} h_{\ell}^{\sigma^{\ell | f}}) = 
 \delta_{r} (\prod_{\ell \notin \tilde \cT }^{\rightarrow} h_{\ell}^{\sigma^{\ell | \cup_{f\in \cH} f} } )
 \prod_{f\in \cH, f\neq r} 
\delta_{f} \Big{(} \tilde h_{l(f,\tilde \cT)}^{\sigma^{l(f,\tilde \cT) | f}} 
\Big{)} \; .
\eea

By a tree change of variables \cite{FreiGurOriti} all the group elements $h_l$ of the lines in the tree 
$l\in \cT$ can be set
to the identity. If $\cH$ is a planar graph, then either $l \in \cT$ or 
$l\in \tilde \cT$ hence $h_l=e ,\; \forall l\in \cH$. This also holds if $\cH$ is a planar graph 
with exactly one external face by setting the latter as root.

\bigskip

\begin{lemma}\label{lem:genujackets}{\bf [Jacket bound]}
Consider a graph $\cG$ with $2p$ vertices and degree $\omega(\cG)$. Then
   \bea
   A^{\text{BO}} (\cG) \le (\lambda\bar\lambda)^p [\delta(e)]^{D-1 -  \frac{2(D-2)}{D!} \omega(\cG)  } \; .
    \eea
\end{lemma}

\noindent{\bf Proof:}
All the group elements $h_l$ of the lines of a tree $l\in \cT$ can be eliminated from the amplitude 
\eqref{eq:ampli} by a tree change of variables \cite{FreiGurOriti}.
Routing the faces of a jacket $\cH$ of $\cG$, eq. \eqref{eq:ampli} writes 
\bea
A^{\text{BO}} (\cG) &=&\frac{(\lambda\bar\lambda)^p}  { [\delta(e)]^{p\frac{ (D-2)(D-1)}{2} } }
\int \prod_{\ell\in \cL_{\cG} \setminus  \tilde \cT } dh_{\ell}  
\prod_{ l\in \tilde \cT } d\tilde h_{l }
\Big{[}  \prod_{f'\notin \cH } 
\delta_{f'}( \dots )  \Big{]} \delta_{r} ( \dots  ) 
\Big{[} \prod_{f\in \cH , f\neq r} \delta_{f} \Big{(} \tilde h_{l(f,\tilde \cT)} \Big{)} \Big{]} 
\nonumber\; , 
\eea
Integrating $\tilde h_{l(f,\tilde \cT)} $ and bounding the remaining delta functions by $\delta(e)$ we obtain
\bea\label{eq:jacketbound}
 A^{\text{BO}} ( \cG ) \le  (\lambda\bar\lambda)^p [\delta(e)]^{ -p\frac{(D-2)(D-1)}{2} + \cF_{\cG} -\cF_{\cH} +1 } \; ,
\eea 
and $A(\cG)$ saturates eq. \eqref{eq:jacketbound} if $\cH$ is planar.
Using \eqref{eq:faces}, eq. \eqref{eq:jacketbound} translates into 
\bea\label{eq:boundfin}
A^{\text{BO}} (\cG) \le  (\lambda\bar\lambda)^p [\delta(e)]^{   D - 1 - \frac{2}{(D-1)!} \sum_{\cJ} g_{\cJ} + 2 g_{\cH}   }
\le (\lambda\bar\lambda)^p [\delta(e)]^{   D - 1 - \frac{2(D-2)}{D!} \sum_{\cJ} g_{\cJ}  }
\; ,
\eea
where for the last inequality we chose $\cH$ with $g_{\cH} = \inf_{\cJ} g_{\cJ}$. 

\qed

\subsection{$k$-Dipoles of degree $0$}

In this section we investigate the behavior of the amplitude of a graph of the BO model under
the contraction of a $k$-Dipole of degree $0$. 

\begin{lemma}\label{lem:ampli}
 If a $k$-Dipole has degree $\omega(d_k)=0$ then 
  \bea
   A^{\text{BO}} (\cG) = (\lambda \bar\lambda) \; \delta(e)^{-(D+1)k +k^2+D } \; A^{\text{BO}} (\cG/d_k) \; .
  \eea  
\end{lemma}

\noindent{\bf Proof:}
 Consider a $k$-Dipole of colors $0,\dots, k-1$. All faces $ij$ such that
$k \le i,j$ touching $v$ are different from the faces $ij$ touching $w$.
Suppose that all lines enter $v$ and exit $w$. 
We denote $h_0, \dots, h_{k-1}$ the group elements of the lines of the $k$-Dipole and
$h_{i;v}, h_{i;w}$, $k\le i$ the ones of the of the lines of color $i$
touching $v$ and $w$ respectively. The rest of the group elements along the 
various faces are generically denoted $f^{ij}_{wv}$ if $i<k$, $k\le j$ and 
$f^{ij}_v$ respectively $f^{ij}_w$ for $k \le i,j$.
 
The contribution of all faces containing $v$ and/or $w$
to the amplitude of $\cG$ is
\bea \label{eq:ampliii}
&& \int \prod_{i=0}^{k-1} dh_{i} \prod_{i=k}^{D} dh_{i;v} \;  dh_{i;w}
 \prod_{0\le i<j < k} \delta^{(ij)} \bigl( h_i^{-1} h_j \bigr) \crcr
&& \prod_{0\le i <k\le j } \delta^{(ij)} \bigl( h_{j;v} h_i^{-1} h_{j;w} f^{ij}_{wv} \bigr) 
 \prod_{k\le i <j} \delta^{(ij)} \bigl( h_{i;v} h_{j;v}^{-1} f_v^{ij} \bigr) 
\; \delta^{(ij)} \bigl(  h_{j;w}^{-1} h_{i;w} f_{w}^{ij}\bigr) \; .
\eea
All $0<h_i<k$ can be integrated explicitly and renaming $h=h_0$ we get 
\bea
 \delta(e)^{\frac{ (k-2) (k-1) }{2} }  \int dh \prod_{i=k}^{D} dh_{i;v} \;  dh_{i;w} 
 \prod_{0\le i <k\le j } \delta^{(ij)} \bigl( h_{j;v} \;  h^{-1}  h_{j;w} f^{ij}_{wv} \bigr) 
 \prod_{k\le i <j} \delta^{(ij)} \bigl( h_{i;v} h_{j;v}^{-1} f_v^{ij} \bigr) 
\; \delta^{(ij)} \bigl(  h_{j;w}^{-1} h_{i;w} f_{w}^{ij}\bigr) \; .
\eea
We change variables to $ h_{j;v}'= h_{j;v} \;  h^{-1}$, $dh_{j;v}'= dh_{j;v} $ for $j\ge k$ and drop the primes.
The integral over $h$ decouples and we get
\bea
 \delta(e)^{\frac{ (k -2 )(k-1) }{2} } \int \prod_{i=k}^{D} dh_{i;v} \;  dh_{i;w} 
 \prod_{0\le i <k\le j } \delta^{(ij)} \bigl( h_{j;v} \;   h_{j;w} f^{ij}_{wv} \bigr) 
 \prod_{k\le i <j} \delta^{(ij)} \bigl( h_{i;v} h_{j;v}^{-1} f_v^{ij} \bigr) 
\; \delta^{(ij)} \bigl( h_{j;w}^{-1} h_{i;w} f_{w}^{ij}\bigr) \; .
 \eea
For $j\ge k$ we change variables to $ h_j = h_{j;v} \;   h_{j;w} $, $dh_{j} = dh_{j;v}$, and we obtain
\bea
&& \delta(e)^{\frac{ (k-2) (k-1) }{2} } \int \prod_{i=k}^{D} dh_{i} \;  dh_{i;w} 
 \prod_{0\le i <k\le j } \delta^{(ij)} \bigl( h_{j} f^{ij}_{wv} \bigr) 
 \prod_{k\le i <j} \delta^{(ij)} \bigl( h_i h_{i;w}^{-1} h_{j;w} h_j^{-1} f_v^{ij} \bigr) 
\; \delta^{(ij)} \bigl(  h_{j;w}^{-1} h_{i;w} f_{w}^{ij}\bigr) \crcr
&&= \delta(e)^{\frac{ (k-2) (k-1) }{2} } \int \prod_{i=k}^{D} dh_{i} \;  dh_{i;w} 
 \prod_{0\le i <k\le j } \delta^{(ij)} \bigl( h_{j} f^{ij}_{wv} \bigr) \;
\prod_{k\le i <j} \delta^{(ij)} \bigl( h_i  f_{w}^{ij} h_j^{-1} f_v^{ij} \bigr) 
 \prod_{k\le i <j} \delta^{(ij)} \bigl(  h_{j;w}^{-1} h_{i;w} f_{w}^{ij}\bigr)  \; .
\eea
We separate the integration variables and write
\bea
\delta(e)^{\frac{ (k -2)(k-1) }{2} } \int dh_k dh_{k;w} \prod_{j=k+1}^D dh_{j} dh_{j;w} 
&& \prod_{0\le i <k\le j } \delta^{(ij)} \bigl( h_{j} f^{ij}_{wv} \bigr) \;
\prod_{k\le i <j} \delta^{(ij)} \bigl( h_i  f_{w}^{ij} h_j^{-1} f_v^{ij} \bigr) \crcr
&& \prod_{k <j} \delta^{(kj)} \bigl(  h_{j;w}^{-1} h_{k;w} f_{w}^{kj}\bigr)  
\prod_{k< i <j} \delta^{(ij)} \bigl(  h_{j;w}^{-1} h_{i;w} f_{w}^{ij}\bigr)  \; .
\eea
All $ h_{j;w}$ with $j>k$ integrate with the faces $kj$, and $h_{k;w}$ drops out. We get
\bea
 \delta(e)^{\frac{ (k-2) (k-1) }{2} } \int \prod_{j=k}^{D} dh_{j} 
&& \prod_{0\le i <k\le j } \delta^{(ij)} \bigl( h_{j} f^{ij}_{wv} \bigr) \;
\prod_{k\le i <j} \delta^{(ij)} \bigl( h_i  f_{w}^{ij} h_j^{-1} f_v^{ij} \bigr) 
\crcr
&& \prod_{k< i <j} \delta^{(ij)} \bigl( (f^{kj}_w)^{-1}  f_{w}^{ki} f_{w}^{ij}\bigr)  \; .
 \eea
The first line reproduces the appropriate factor contributing to the amplitude of $\cG/d_k$.
The second line counts $\frac{1}{2} ( D-k )(D-k-1) $ terms. 
All the relations corresponding to the closed faces
of $\cB^{\widehat{0} \dots \widehat{k-1} }_{(\alpha)} \setminus w$ (and any of its jackets)
appear in the amplitude of $\cG/d_k$. As $d_k$ has degree $0$,
all the jackets $\cJ^{\widehat{0} \dots \widehat{k-1}}_{(\alpha)} $ are planar hence
$\cJ^{\widehat{0} \dots \widehat{k-1}} \setminus w$ is a planar graph with 
one external face. Setting $h_l = e$ for some tree  
in $\cJ^{\widehat{0} \dots \widehat{k-1} }_{(\alpha)} \setminus w$ and routing the faces
leads to $ f^{ij}_{w}=e$. Taking into account that $\cG/d_k$ has two vertices less that $\cG$ we have
\bea
 A^{\text{BO}} (\cG) = \lambda \bar\lambda \; \delta(e)^{-\frac{(D-2)(D-1)}{2} + \frac{(k-2)(k-1)}{2} + \frac{(D-k)(D-k-1)}{2}}
 A^{\text{BO}} (\cG/d_k) \; .
\eea

\qed

\subsection{Topological $1/N$ expansion}

The power counting of graphs of the topological model is invariant under contractions of
$1$-Dipoles of degree $0$. Following the procedure in section \ref{sec:topo1/N} we perform a 
topological bubble routing of any graph $\cG$ to a topological core graph $\cG^{(t)}_p$. 
Combining eq. \eqref{eq:degtopol} with lemma \ref{lem:genujackets}
yields 

\begin{lemma}\label{lem:boundcore}
    The amplitude of a topological core graph $\cG^{(t)}_p$ with $2p$ vertices is bounded by
    \bea
    A^{\text{BO}} (\cG^{(t)}_p) \le (\lambda\bar\lambda)^p \delta(e)^{ D-1 - \frac{2(D-2)}{D!} (p-1) } \; .
    \eea
\end{lemma}

The free energy of the Boulatov Ooguri model writes in terms of topological core equivalence classes as
\bea\label{eq:topotopo}
 F^{\text{BO}}(\lambda, \bar\lambda) = \sum_{p=1}^{\infty} \sum_{[\cG^{(t)}_p] } 
C^{[\cG^{(t)}_p]}(\lambda,\bar\lambda) A^{\text{BO}} ([\cG^{(t)}_p]) \; .
\eea

The equation \eqref{eq:topotopo} and lemma \ref{lem:boundcore} encode the $1/N$ expansion of the Boulatov Ooguri 
model in arbitrary dimension. To determine the 
series up to order $\delta(e)^{D-1-\alpha}$, one must list all core graphs up to 
$p_{max} = 1 + \frac{D!}{2(D-2)} \alpha$ and classify them in topological core equivalence classes. 

The topological expansions \eqref{eq:topoiid} and \eqref{eq:topotopo} group together only graph with both 
the same topology and amplitude. However the scaling with $\delta(e)$ only separates graphs of different amplitudes,
 as captured by the combinatorial expansion of the i.i.d. model \eqref{eq:free}.
Classifying further the graphs with the same amplitude into {\it topological} core equivalence classes is a choice.
This leads to a well defined series one can compute, but the topological expansion has a certain degree 
of redundancy
\begin{itemize}
\item  There exist graphs with $A(\cG) = A(\cG')$ but $\cG \nsim^{(t)} \cG'$. Although the scaling with $\delta(e)$
       does not distinguish two such graphs, we obviously group them in different terms in \eqref{eq:topoiid} and 
       \eqref{eq:topotopo}.
\item  Any topology appears at arbitrary order. For all topological core graphs $\cG^{(t)}_p$ one obtains 
       topologically equivalent core graphs (but smaller in power counting) by  creating an arbitrary number of 
       $k$-Dipoles of degree 0 for $k\ge2$.
\item  At the same order $p$, the same topology can appear in several distinct classes 
       ($\cG^{(t)}_p \sim^{(t)} \cG^{(t)}_p{}'$ but $A^{\cG^{(t)}_p}\neq A^{\cG^{(t)}_p{}'}$). 
\item  There exist topological core classes $[\cG^{(t)}_p]$ and $[\cG^{(t)}_q]$ at different orders $p\neq q$, 
        with both $\cG^{(t)}_p \sim \cG^{(t)}_q$ and $A^{\cG^{(t)}_p}=A^{\cG^{(t)}_q}$.
       (for instance if $\cG^{(t)}_q$ can be obtained from $\cG^{(t)}_p$ by adding $1$-Dipoles of degree higher than 
        $\omega(d_1)\ge 1$ which separate spheres\footnote{Note that, as the degree of the $D$-bubbles is additive under 
        $1$-Dipole creations, $q\le p+ \omega(\cG_p)$.}). 
\end{itemize}

As the topology in higher dimensions is very involved, it is to be expected that the final $1/N$ 
expansion is somewhat difficult to handle. Counting all graphs which route to a given core 
graph is relatively straightforward (by performing all possible creations of $1$-Dipoles of degree $0$). However
one must pay attention to double counting problems when grouping together the graphs routing to a 
topological core equivalence class. Fortunately, at low orders, the topological core equivalence classes 
have a single representative, and the overcounting problem is absent.

A bubble $\cB^{\widehat{i}}_{(\rho)}$ of a $3+1$ colored graph $\cG$ represents a sphere 
if and only if $\omega( \cB^{\widehat{i}}_{(\rho)})=0$ hence a $1$-Dipole $d_1$ separates a sphere
if and only if  $\omega(d_1)=0$. This eliminates most of the redundancies of the $1/N$ expansion. 

\bigskip

\begin{figure}[htb]
\begin{center}
 \includegraphics[width=1.5cm]{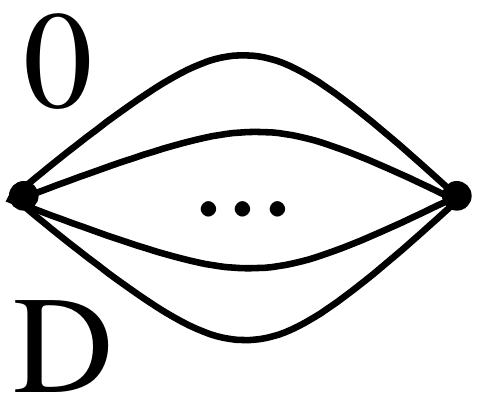} \hspace{1cm}
 \includegraphics[width=2cm]{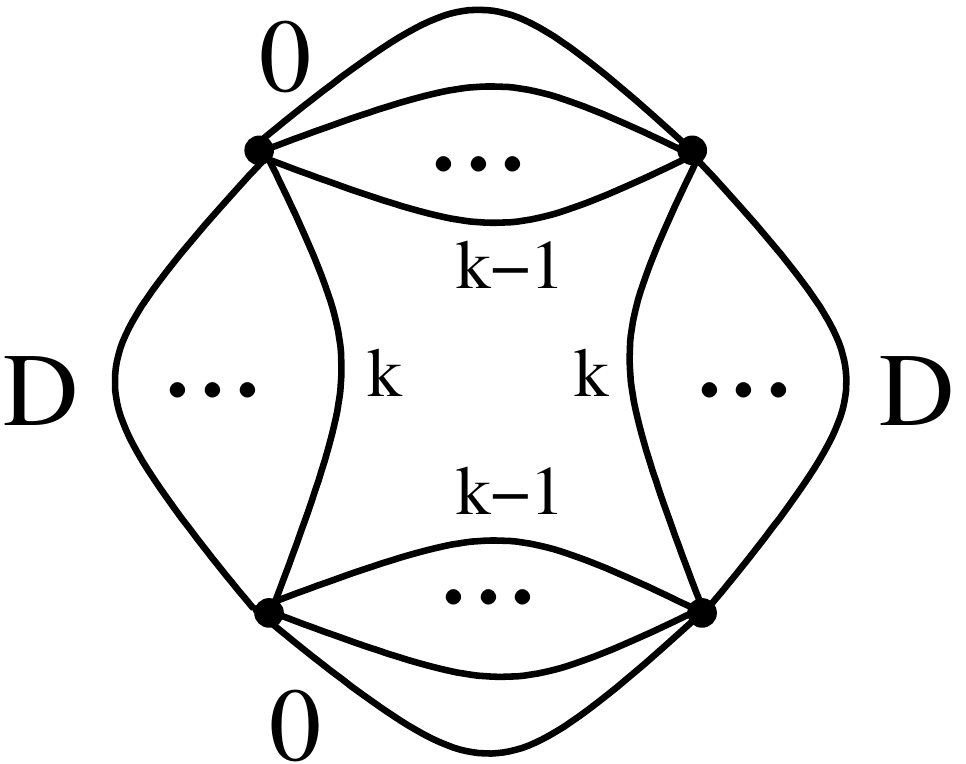}
\caption{Core graphs at $p=1$ and $p=2$.}
\label{fig:t1t2}
\end{center}
\end{figure}

\section{The first terms of the $1/N$ expansion}\label{sec:firstterms}

All combinatorial core graphs are also topological core graphs. 
The topological core graphs up to $p=2$ are represented in figure \ref{fig:t1t2}.
As they only have one $D$-bubble $\cB^{\widehat{i}}$ they are also combinatorial
core graphs.

At $p=1$ we have a unique topological/combinatorial core graph, denoted $\cG^{(c)}_1=\cG^{(t)}_1$. It is 
the unique topological/combinatorial core 
graph of degree $0$, thus it represents a sphere. As the degree of 
a graph is invariant under arbitrary $1$-Dipole contractions, the 
combinatorial class $[\cG^{(c)}_1]$ is identical with the topological class $[\cG^{(t)}_1]$. 

At $p=2$ we have the topological/combinatorial core graphs $\cG^{(c)}_{2;k}=\cG^{(t)}_{2;k}$, 
with $2\le k \le [\frac{D+1}{2}]$. This time the combinatorial and topological classes
$[\cG^{(c)}_{2;k}]$, $[\cG^{(t)}_{2;k}]$ are different. 
All these core graphs are dual to spheres $S^D$ ($\cG^{(t)}_1\sim^{(t)} \cG^{(t)}_{2;k}$). 
Their amplitudes are
\bea
&& A^{\text{i.i.d}}(\cG_1) = \delta(e)^{D} \qquad A^{\text{i.i.d}} (\cG_{2;k}) = \delta(e)^{D-(D+1)k+k^2+D} \; ,
  \crcr
&& A^{\text{BO}} (\cG_1) = \delta(e)^{D-1} \qquad A^{\text{BO}} (\cG_{2;k}) = \delta(e)^{D-1-(D+1)k+k^2+D} \; .
\eea

The various $1/N$ expansions of the free energy of the models we have considered in arbitrary dimensions write
\bea\label{eq:final}
&& F^{\text{i.i.d.}}(\lambda,\bar\lambda) = C^{[\cG^{(c)}_1]}(\lambda,\bar\lambda) \; \delta(e)^{D} + 
   \sum_{k=2}^{ [\frac{D+1}{2}] } C^{[\cG^{(c)}_{2;k}]}(\lambda,\bar\lambda)  \; \delta(e)^{D-(D+1)k+k^2+D}
   +O \bigl( \delta(e)^{D - 2 } \bigr) \; ,
\crcr
&& F^{\text{i.i.d.}}(\lambda,\bar\lambda) = C^{[\cG^{(t)}_1]}(\lambda,\bar\lambda) \; \delta(e)^{D} + 
   \sum_{k=2}^{ [\frac{D+1}{2}] } C^{[\cG^{(t)}_{2;k}]}(\lambda,\bar\lambda)  \; \delta(e)^{D-(D+1)k+k^2+D}
   +O \bigl( \delta(e)^{D - \frac{2}{(D-1)!} 2 } \bigr) \; ,
\crcr
&& F^{\text{BO}}(\lambda,\bar\lambda) = C^{[\cG^{(t)}_1]}(\lambda,\bar\lambda) \; \delta(e)^{D-1} + 
   \sum_{k=2}^{ [\frac{D+1}{2}] } C^{[\cG^{(t)}_{2;k}]}(\lambda,\bar\lambda)  \; \delta(e)^{D-1-(D+1)k+k^2+D}
   +O \bigl( \delta(e)^{D-1 - \frac{2 (D-2) }{D!} 2 } \bigr) \; .
\eea

Note that for the i.i.d. model the combinatorial expansion has a better a priori bound on the reminder terms 
than the topological expansion. This should come as no surprise, as the topological classes 
$[\cG^{(t)}_{2;k}]$ contain fewer graphs than the combinatorial classes $[\cG^{(c)}_{2;k}]$.
The a priori bound on the reminder terms is not very tight. 
The first explicit correction w.r.t the dominant behavior in eq. \eqref{eq:final} comes from 
$\cG_{2,2}$. To ensure that all terms with amplitude larger or equal than $\cG_{2,2}$ have 
been taken into account,  one needs to check topological core graphs up to 
$p_{\text{max}}^{\text{i.i.d.}}=1 +\frac{(D-1)!}{2} (D-2)$ and $p^{\text{BO}}_{\text{max} }=1+\frac{D!}{2}$
in the topological expansions combinatorial core graphs up to $p_{\text{max}}=D-1$ in the combinatorial 
expansion of the i.i.d. model.

\section*{Acknowledgements}

The author would like to thank Vincent Rivasseau for the numerous 
discussions on the topic of the $1/N$ expansion which have been at the foundation of this work.

Research at Perimeter Institute is supported by the Government of Canada through Industry 
Canada and by the Province of Ontario through the Ministry of Research and Innovation.

\end{document}